\documentclass[aps,prl,superscriptaddress,amsmath,amssymb,floatfix,twocolumn]{revtex4-1}
\usepackage{times}
\usepackage{graphicx,graphics,color,epsfig}% Include figure files
\usepackage{subfigure}
\usepackage{color}

\begin{document}

\title{Local Electronic Structure around a Single Impurity
as a Test of Pairing Symmetry in  (K,Tl)Fe$_{x}$Se$_{\rm 2}$ Superconductors}

\author{Jian-Xin Zhu}
\email[To whom correspondence should be addressed. \\ Electronic address: ]{jxzhu@lanl.gov}
\homepage{http://theory.lanl.gov}
\affiliation{Theoretical Division, Los Alamos National Laboratory,
Los Alamos, New Mexico 87545}

\author{Rong Yu}
\affiliation{Department of Physics \& Astronomy, Rice University, Houston, Texas 77005}

\author{A. V. Balatsky}
\affiliation{Theoretical Division, Los Alamos National Laboratory,
Los Alamos, New Mexico 87545}
\affiliation{Center for Integrated Nanotechnologies, Los Alamos National Laboratory, Los Alamos,
New Mexico 87545}

\author{Qimiao Si}
\affiliation{Department of Physics \& Astronomy, Rice University, Houston, Texas 77005}

\begin{abstract}
We have studied the effect of a single nonmagnetic impurity in the recently
discovered (K,Tl)Fe$_x$Se$_2$ superconductors,
within both a toy two-band model and a more realistic five-band model.
We have found that, out of five types of pairing
symmetry under consideration, only the $d_{x^2-y^2}$-wave pairing gives rise to
impurity resonance states. The intra-gap states have energies far away from the Fermi energy.
The existence of these intra-gap states is robust against the presence or absence of inter-band scattering.
However, the inter-band scattering does tune the relative distribution of local density of states at the resonance
states. All these features can readily be accessed by STM experiments,
and are proposed as a means to test pairing symmetry of  the new superconductors.
 \end{abstract}
%74.25.Jb	Electronic structure (photoemission, etc.)
%74.20.Pq	Electronic structure calculations (for methods of electronic structure calculations, see 71.15.-m)
%74.20.-z	         Theories and models of superconducting state
%74.62.En	Effects of disorder
%74.55.+v	Tunneling phenomena: single particle tunneling and STM
\maketitle

{\it Introduction.~} The very recent discovery of high-$T_c$ (above 30 K) superconductivity in
AFe$_{x}$Se$_2$ (A= K, Tl, Cs)~\cite{JGuo:2010,AKMaziopa:2011,MFang:2010} has generated a new wave of excitement in the field of iron-based superconductors.
The new family of compounds is unique in several regards.
(i) the superconducting transition temperature is about four 
times that of FeSe under ambient pressure (which is roughly 8 K)~\cite{FHsu:2008}; (ii) in contrast to the bad metal behavior in other iron-based parent compounds, the Fe-deficient compound ($x \leq 1.6$) is insulating~\cite{MFang:2010,DMWang}, raising the interest in the possibility of Mott insulating state~\cite{XWYan:2010a, CCao:2011,RYu:2011,YZhou:2011}
induced by patterned Fe-vacancies~\cite{ZWang:2011,LHaggstrom:1991};
%(i) Like FeSe, it does not contain any toxic arsenic element, but the superconducting
%transition temperature
%is about four times that of FeSe under ambient pressure (which is roughly 8 K)~\cite{FHsu:2008};
%(ii) The Fe-deficient compound ($x \leq 1.6$) shows insulating behavior~\cite{MFang:2010,DMWang}. This is
% in contrast to other iron-based parent compounds, which are bad metals,
% raising the interest in the possibility of Mott insulating
% state~\cite{XWYan:2010a, CCao:2011,RYu:2011,YZhou:2011}
%induced by patterned
%Fe-vacancies~\cite{ZWang:2011,LHaggstrom:1991};
(iii) The end members, TlFe$_2$Se$_2$ and KFe$_2$Se$_2$, are heavily electron doped (0.5 electron/Fe) relative to other iron-based superconductors (such as LaOFeAs, BaFe$_2$Se$_2$, FeSe etc.). Band structure calculations~\cite{LZhang:2009,XWYan:2010b,CCao:2010,IShein:2010,INekrasov:2011}
 on these end compounds show only electron pockets, primarily located
  around the $M$ point of the Brillouin zone (BZ) defined for a simple tetragonal structure.
Angle-resolved photoemission spectroscopy (ARPES) measurements
observed these electron-like pockets around the $M$ points, and showed
no hole-like pockets~\cite{TQian:2010,YZhang:2010} but very weak
electron-like pockets~\cite{DMou:2011,XPWang:2011,LZhao:2011} near the zone center $\Gamma$.

The superconducting pairing symmetry of iron-based superconductors has been extensively discussed from both weak-coupling and strong-coupling approaches. Within weak-coupling approaches,
the usual argument for the popular $s_{\pm}$ pairing symmetry relies crucially
on the existence of $\Gamma$-centered hole pockets and $M$-centered electron pockets,
which was the case in previously studied
iron-based superconductors. By contrast, strong-coupling approaches invoke pairing driven by
$J_1$-$J_2$ exchange interactions.  The absence of $\Gamma$-centered hole pockets, therefore, provides an opportunity to elucidate the mechanism for iron-based superconductivity.
Recent calculations have predicted that the superconducting state could have $d_{x^2-y^2}$-wave~\cite{FWang:2011,TMaier:2011,TDas:2011}, $s$-wave symmetry~\cite{YZhou:2011},
or  a  mixture of
$s_{x^2y^2}$- and $d_{x^2-y^2}$-wave pairing states in the magnetic
frustration regime~\cite{RYu:2011b}.
All these scenarios lead to nodeless superconducting gap structure, which is
 in agreement with the ARPES observations  and 
 other experiments~\cite{LMa:2011,BZeng:2011}.
Because of the particular Fermi surfaces, conventional phase-sensitive measurements
cannot be readily applied to differentiate the different pairing states.

In this Letter, we
propose to use local electronic structure around a single nonmagnetic impurity to probe the pairing symmetry in  (K,Tl)Fe$_{x}$Se$_{2}$ superconductors. Such properties have proved to be fruitful in
 identifying  the unconventional pairing states of different classes of
  superconductors~\cite{AVBalatsky:2006}. The effort
has been complicated~\cite{TZhou:2009,WTsai:2009} in the previously
 studied Fe-based superconductors due to the existence of
both hole-like and electron-like  Fermi surface pockets.
Because the Fermi surface of the new (K,Tl)Fe$_{x}$Se$_{2}$ compounds comprise
 small pockets of only  one type of carriers,
this kind of study might be very promising.  We study the problem within both
a two-band toy model and an effective model including detailed band-structure from LDA calculations.
Our results within a $T$-matrix approximation show that
the intragap impurity-induced bound state only exists when the pairing symmetry is of $d_{x^2-y^2}$.
We also find that the bound-state peak in the local density of states (LDOS) occurs at a non-zero energy even in the
unitary limit, in contrast to the situation in high-$T_c$ cuprates.  Our prediction can be directly tested
by the scanning tunneling microscopy experiments on these new compounds.

{\it Local impurity effects in a toy two-band model.~} We start with a two-dimensional (2D) lattice formed
by Fe atoms, and consider a two-band toy model with the energy dispersion:
\begin{equation}
\varepsilon_{m,\mathbf{k}} = -4D_{m} \cos k_x \cos k_y + \epsilon_m\;,
\label{EQ:DISP}
\end{equation}
for each band $m$ ($=1,2$) with the band width $8|D_m|$ and band center-of-gravity $\epsilon_m$.
In the superconducting state, the bare Green's function within the Nambu space is given by
\begin{equation}
\hat{\mathcal{G}}_{0}^{-1}(\mathbf{k},i\omega_n) =  i\omega_{n} \hat{1} -
\left( \begin{array}{cccc}
\varepsilon_{1,\mathbf{k}} & \Delta_{1,\mathbf{k}}  & 0   & 0  \\
\Delta_{1,\mathbf{k}}  & -\varepsilon_{1,\mathbf{k}} & 0  & 0 \\
0 & 0  &  \varepsilon_{2,\mathbf{k}} & \Delta_{2,\mathbf{k}}  \\
0 & 0 & \Delta_{2,\mathbf{k}}  & -\varepsilon_{2,\mathbf{k}}
\end{array} \right)\;,
\end{equation}
where $\omega_n=(2n+1)\pi T$ ($n$ integer) is the Matsubara frequency for fermions.
The superconducting gap function is described by $\Delta_{m,\mathbf{k}}$. Hereafter, we consider
only spin singlet pairing.  When an alien atom is substituted for Fe, it plays the role of a short-ranged potential scatter. In the single-site impurity approximation, the Green's function is
 given with the $T$-matrix approximation by:
\begin{equation}
\hat{\mathcal{G}}(i,j;i\omega_n) = \hat{\mathcal{G}}_0(i,j;i\omega_n) + \hat{\mathcal{G}}_0(i,0;i\omega_n)\hat{T}\hat{\mathcal{G}}_0(0,j;i\omega_n)\;.
\end{equation}
Here the $T$-matrix can be expressed as $\hat{T}(i\omega_n) = \hat{U}[\hat{1} - \hat{\mathcal{G}}_0(0,0;i\omega_n)\hat{U}]^{-1}$. The potential scattering matrix takes the following structure
\begin{equation}
\hat{U} = \left( \begin{array}{cccc}
u  & 0  & v   & 0  \\
0  & -u  & 0  & -v \\
v   &  0  &  u  & 0 \\
0   & -v & 0  & -u
\end{array} \right)\;,
\end{equation}
where $u$ and $v$ are the strength of the intra- and inter-band scattering potential.
The bare real-space Green's function can be obtained from the Fourier transform:
$\hat{\mathcal{G}}_0(i,j;i\omega_n) = \frac{1}{N_{L}}\sum_{\mathbf{k}} \hat{\mathcal{G}}_0(\mathbf{k};i\omega_n)
e^{i\mathbf{k}\cdot(\mathbf{r}_{i}-\mathbf{r}_j)}$, where $N_L$ is the number of lattice sites.
The LDOS can be evaluated as $\rho_{i}(\omega) = -\frac{2}{\pi}\text{Im}\{\text{Tr}[\mathcal{G}(i,i;\omega+i\eta)]\}$ with $\eta$ being the intrinsic lifetime broadening parameter. This quantity is proportional
to the local differential tunneling conductance as measured in STM experiments.
The above scheme is sufficiently general to capture essential
properties of the single impurity scattering in a two-band superconductor.

\begin{figure}[t]
%\centering\includegraphics[scale=0.28]{fig1.eps}
\centerline{\psfig{figure=fig1a.eps,width=4cm,angle=0}
\psfig{figure=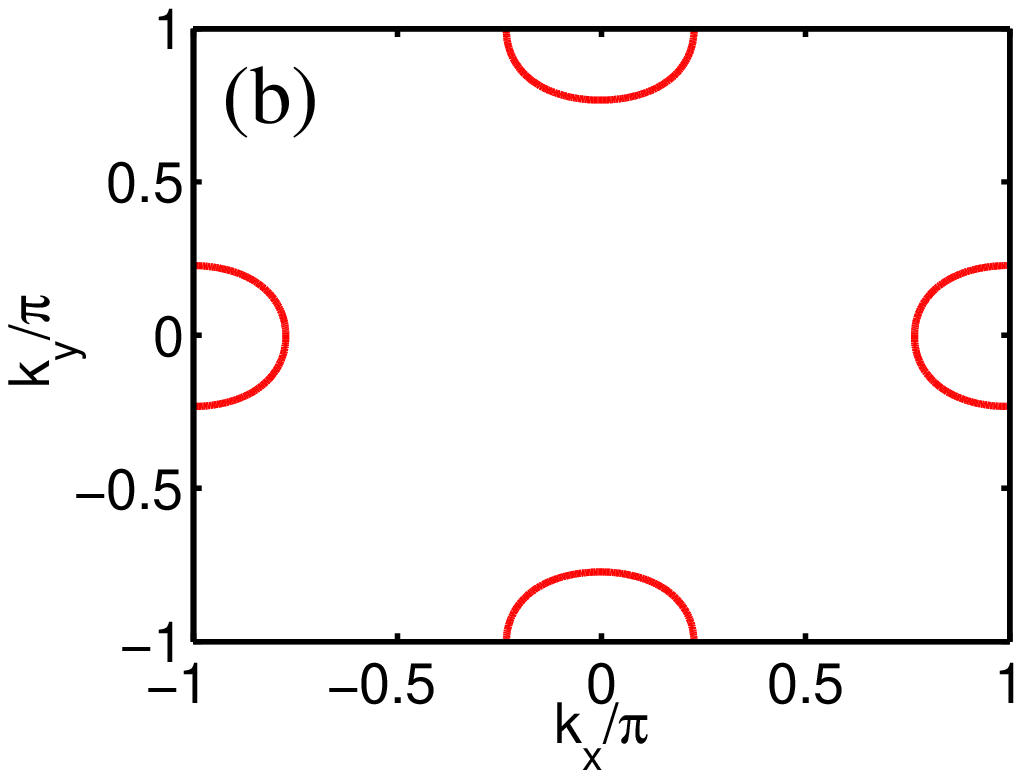,width=4cm,angle=0}}
\caption{(Color online) Energy dispersion along the cut at $k_y=0$ (a) 
and the Fermi surface topology (b) in the large one Fe per cell BZ based 
on a toy two-band model with model parameters given in the text.
%Energy dispersion along the cut at $k_y=0$ in the large one Fe per cell Brillouin zone (a)
%and the Fermi surface topology in the BZ (b) based on a toy two-band model. Here $D_1=-1$, $D_2=-0.5$,
%$\epsilon_1=3.0$, and $\epsilon_2=-2.5$ are chosen so that there are only electron Fermi surface pockets
%in the BZ.
}
\label{FIG:DISP}
\end{figure}

 To mimic the electron pockets of (K,Tl)Fe$_{x}$Se$_{2}$ superconductors
within the present toy model, we take $D_1=-1$, $D_2=-0.5$, $\epsilon_1=3$,
 and $\epsilon_2=-2.5$ in the two-band energy dispersion as given by Eq.~(\ref{EQ:DISP}), which leads to an electron occupation of 2.17 with the zero value of chemical potential. Unless
 specified otherwise, the energy is measured in units of $\vert D_1 \vert$.
 Figure~\ref{FIG:DISP} shows the energy dispersion along the cut with $k_y=0$ and the Fermi surface in the one Fe per unit cell BZ.

We now turn to the response of the local electronic structure to the single impurity scattering in the superconducting state with various pairing symmetry. The considered pairing symmetry includes
conventional momentum-independent $s$-wave channel with $\Delta_{m,\mathbf{k}}=\Delta_m^{(0)}$ (s0),
the extended $s$-wave channel with $\Delta_{m,\mathbf{k}} = \frac{\Delta_m^{(0)}}{2}(\cos k_x + \cos k_y)$ (s1)
as well as $\Delta_{m,\mathbf{k}} = \Delta_m^{(0)} \cos k_x  \cos k_y$ (s2), and the $d$-wave channel
with $\Delta_{m,\mathbf{k}} = \frac{\Delta_m^{(0)}}{2}(\cos k_x - \cos k_y)$ (d1) as well as
$\Delta_{m,\mathbf{k}} = \Delta_m^{(0)} \sin k_x  \sin k_y$ (d2).
For the superconducting gap (s0), there is no sign change within each electron pocket or across
the pockets along the zone boundary.
When the radius of the electron pockets smaller than $\pi/2$, this is also the case for
the gap structure (s2). However, the gap structure (s1) and (d2) have the sign change within each electron pocket. The gap structure (d1) has no sign change within each electron pocket but changes
sign across  two neighboring electron pockets. The various types of pairing symmetry have been
considered intensively in the studies of early Fe-arsenic based superconductors.

\begin{figure}[t]
\centering\includegraphics[%scale=3.0
width=0.95\linewidth
]{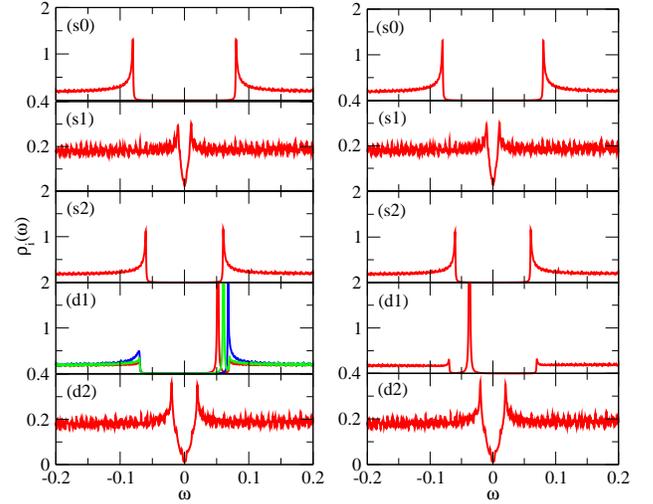}
\caption{(Color online) 
LDOS measured at the first nearest neighboring sites to the single impurity site
with only intra-band scattering (left column) and with both intra- and inter-band scattering (right column).
Each column is divided into panels for different types of pairing symmetry: (s0), (s1), (s2), (d1), and (d2). Except for the panel for (d1) in the left column, where $u=1, \;5, \;100$
is taken, the scattering strength is taken as $u=100$ and $v=0$ for the left column or $u=v=100$ for the
right column. The superconducting pairing potential is taken as $\Delta_{1}^{(0)}=-\Delta_{2}^{(0)}=0.08$.
The intrinsic broadening parameter $\eta = 5\times 10^{-4}$.
}
\label{FIG:LDOS_TOY}
\end{figure}

In our numerical simulation, we have taken the gap functions in both bands to have
same magnitude but out of phase with each other.
We notice that since there are only electron pockets from band 1 while the band 2 is below the
Fermi energy, it is not important whether there is a gap on the second band as well as its relative phase
with that in the first band.
Figure~\ref{FIG:LDOS_TOY} shows the LDOS of quasiparticles at a site nearest neighboring to the impurity site, which we assume to be located at the center of the square lattice.  The left column of the figure
is for the case without inter-band impurity scattering (i.e., $v=0$) while the right column is for
the case with inter-band impurity scattering $v=u$.
The coherent peak structure related to the system without impurity substitution nicely reflects
the momentum dependence of the gap structure as just analyzed above.
For comparison, we have taken the same magnitude of the pairing potential $\Delta_m^{(0)}$
for all types of pairing symmetry.  Wide quasiparticle gap opens
for the (s0), (s2) and (d1) pairing states, with their well-like shape being similar to what happens
in conventional
$s$-wave superconductors and compatible to the nodeless gap structure
as revealed in the ARPES experiments.  For the (s1) and (d2) pairing symmetry, a $V$-shape DOS
is exhibited reflecting the existence of nodal structure in these two types of pairing symmetry.
Noticeably, the quasiparticle gap size as measured by the distance between two coherent peaks in the
DOS is much smaller than that for (s0), (s2), and (d1) pairing state.  Given the same magnitude
of the pair potential, it has already suggested that the superconducting states with (s1) and (d2) pairing
symmetries do not have enough condensation energy; this is consistent with the aforementioned
microscopic calculations.

We now turn to the the impurity induced
resonance peaks when the impurity scattering is  in the unitary limit.
 We find that such peaks arise only when the superconducting state has
  the (d1) (i.e.,  $d_{x^2-y^2}$-wave)
pairing symmetry.
To make sure that the intra-gap resonance peak originates from the impurity scattering,
we have also calculated the LDOS by varying the impurity scattering potential strength.
As shown in the panel (d1) of the left column, the resonance peak moves toward the coherent gap edge
with decreasing potential strength.  It is interesting to notice that even in the unitary limit, the
intra-gap resonance peak is located far away from the Fermi energy (at $\omega=0$), in contrast to the
case in high-$T_c$ cuprates, where the impurity resonance peak is located very close to the Fermi
energy~\cite{AVBalatsky:2006}.  Two remarks are in order.
%This new phenomenon is related to the two aspects.
Firstly, the electron pockets here are small in size,
implying a strong violation of particle-hole symmetry when the Fermi energy is located near the band bottom.
This aspect has been tested to be true for the $d$-wave superconductor as relevant to high-$T_c$
cuprates~\cite{AVBalatsky:2006}. Secondly, although there is a sign change of the gap structure across two
neighboring electron pockets along the zone boundary, no nodal quasiparticles (with zero energy)
are allowed in the (d1) pairing state. Inclusion of the interband impurity scattering (see the right column
of the figure) does not change the above conclusion except that the intensity peak is tuned to the opposite
side with respect to the Fermi energy for a given measure point of the LDOS.

\begin{figure}[t!]
\centering\includegraphics[%scale=3.0
width=0.8\linewidth]{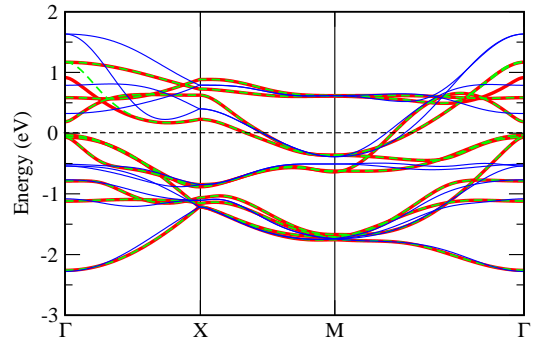}
%\vskip -1.5cm
%\centering\includegraphics[%scale=3.0
%width=1.0\linewidth]{fig3cd.eps}
%\vskip -0.5cm
\caption{(Color online)
The five-orbital tight-binding fit (thin solid blue lines) of the ten-orbital Wannier fit (thin dashed green
lines) to the paramagnetic DFT band structure (thick solid red lines). The splitting of bands near the Fermi
energy at $\Gamma$ point has been enlarged to guarantee the removal of hole pockets at lower electron
doping.
}
\label{FIG:LDA}
\end{figure}

{\it Local impurity effects in an effective low-energy model with realistic band structure.~} To include the realistic
band structure at low energies into the study, we have performed band structure calculations for KFe$_2$Se$_2$ based on the local density approximation.  The full-potential linearized augmented plane wave (FP-LAPW)
method as implemented in the WIEN2K code~\cite{PBlaha:2001} is used. We then follow the procedure
suggested by Graser {\em et al.}~\cite{SGraser:2010} to fit the Wannierized bands~\cite{NMarzari:1997,ISouza:2001}  with a five-orbital tight-banding Hamiltonian~\cite{SGraser:2009},  unfolding the small two Fe per unit cell BZ to a large one Fe per unit cell BZ. In this procedure, an interface~\cite{JKunes:2010} between the WIEN2k code and the wannier90 code~\cite{AMostofi:2008} is also employed.
%The construction of  an effective model based on
%one Fe per unit cell for a body-centered tetragonal (BCT) structure, which is the case for  KFe$_2$Se$_2$ type of systems, is nontrivial. We follow the prescription proposed by Graser {\em et al.}
% in their study of BaFe$_2$As$_2$ compound.
%We first  carry out the calculations with the experimental
%lattice parameters for the BCT unit cell corresponding to the $I4/mmm$ symmetry of the crystal
%but plot the bands along the high symmetry lines of a corresponding simple tetragonal unit cell.
%We use an interface~\cite{JKunes:2010} between the WIEN2k code and the wannier90~\cite{AMostofi:2008}
%to project the bands in the vicinity of the Fermi energy onto the Fe 3$d$ orbitals.  In the spirit
%of maximally localized Wannier functions~\cite{NMarzari:1997,ISouza:2001},
%the bands are disentangled by minimizing
%the spread of the Wannier functions.
%Finally, we fit the Wannier bands with a five-orbital tight-binding Hamiltonian~\cite{SGraser:2009}, unfolding %the small two Fe per unit cell BZ to a large one Fe per unit cell BZ.
As shown in Fig.~\ref{FIG:LDA},
the tight-binding band structure reproduces the LDA band structure pretty accurately, and agrees reasonably well with those results reported earlier~\cite{TMaier:2011}.
Considering  the compound
K$_{0.8}$Fe$_{1.7}$Se$_2$ on which the recent ARPES experiments were performed~\cite{TQian:2010},
the electron doping is reduced from 0.5 electron per Fe to 0.1 electron per Fe.
Correspondingly, as in Ref.~\onlinecite{TMaier:2011}, we have artificially adjusted a few tight-binding fitting parameters to ensure only electron pockets
existing in  the  BZ  as the chemical potential  $\mu=-0.28\;\text{eV}$ is introduced to produce a correct filling factor.

\begin{figure}[t!]
\centering\includegraphics[%scale=3.0
width=0.7\linewidth]{fig4a.eps}
\vskip -0.2cm
\centering\includegraphics[%scale=3.0
width=1.0\linewidth
]{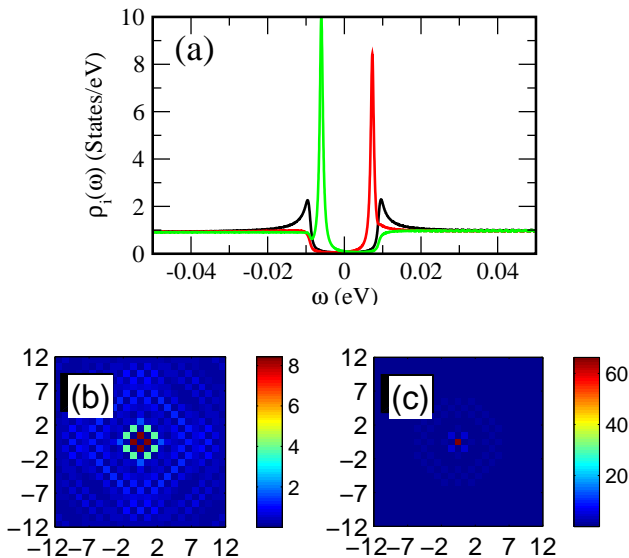}
\caption{(Color online) LDOS at the 1st nearest neighboring site to the single
impurity site (a) and the LDOS imaging (b-c) in the superconductor with (d1) pairing symmetry.
In panel (a), the bare DOS (black line) and LDOS without   ($u=100\;\text{eV}$ and $v=0$) (red line) and with ($u=v=100\;\text{eV}$) inter-band scattering are shown. The resonance peak $E_r$ is at $7.3 \;\text{meV}$ for the former while $6\;\text{meV}$ for the latter when the pair
potential  $\Delta_{m}^{(0)}=10\; \text{meV}$ is chosen.  The LDOS imaging is taken
at $\omega=E_r$ without (panel (b)) and with (panel (c)) inter-band scattering with an intrinsic broadening parameter $\eta = 0.5\; \text{meV}$.
}
\label{FIG:LDA-LDOS}
\end{figure}

The tight-binding Hamiltonian as obtained with the above procedure makes it theoretically possible to
study the low energy quasiparticle properties, where the experimentally extracted
superconducting gap is at the order of 10 meV.  Numerically, we diagonalize the five-orbital tight-binding
model with an extremely large number of $\mathbf{k}$ points in the large one Fe per cell BZ.
Out of the 5 bands from the diagonalization, the two bands with the one cutting the Fermi energy
and the other located immediately below the Fermi energy are picked out for the calculations
of local electronic structure around a single nonmagnetic impurity in the superconducting state.
Guided by the findings from the toy model studied above, we consider here only the superconducting
state with (d1), that is, $d_{x^2-y^2}$-wave, pairing symmetry.  Figure~\ref{FIG:LDA-LDOS}(a) shows the
density of states in the system without the impurity and the LDOS at the site nearest neighboring
to the impurity site. As shown in the toy model, the bare density of states exhibits a well-like gap feature
around the Fermi energy, indicating a nodeless pairing state. Also in the presence of a singe impurity scattering
in the unitary limit, intragap resonance peak appears in the LDOS.
The existence of impurity resonance state is also robust against the inter-band impurity scattering.
In comparison to the case of only intra-band impurity potential scattering, the location of the resonance
energy and its relative intensity on a given measure site might be adjusted by the relative strength of the inter-band scattering.
For a given configuration of impurity scattering, the relative intensity
of the resonance state depends on the location of the measure site.
To better understand the profile of the impurity-induced resonance states, we have also calculated
the spatial dependence of the LDOS at the resonance energy, corresponding to the peak
locations in Fig.~\ref{FIG:LDA-LDOS}(a). The results are shown in Fig.~\ref{FIG:LDA-LDOS}(b-c).
For the case without inter-band scattering, the LDOS intensity is vanishingly small at the impurity site
regardless of whether the selected energy corresponding to the peak at the side above or below the Fermi energy (cf. Fig.~\ref{FIG:LDA-LDOS}(b)). For $\omega=E_r$, the strongest peak intensity is located at the
1st nearest neighboring sites while for $\omega=-E_r$ (not shown), the strongest peak intensity is obtained at the 3rd n.n.  sites.  In addition, the squarish ripples of LDOS distribution is visible for $\omega=\pm E_r$.
For the case with the inter-band scattering strength the same as the intra-band one,
the profile of the LDOS distribution  for  $\omega=-E_r$ is very similar to (cf. Fig.~\ref{FIG:LDA-LDOS}(b)).
However, when $\omega=E_r$ the strongest peak intensity
is located at the impurity site itself  (cf. Fig.~\ref{FIG:LDA-LDOS}(c)) and the LDOS distribution is much more localized than the case with the inter-band scattering.
This unique property, associated with the inter-band scattering,  might enable the usage of the STM technique
to characterize the relative strength of the inter-band scattering.

{\em Conclusion.~} We have studied the effect of a single nonmagnetic impurity in the more recently
discovered (Tl,K)Fe$_2$Se$_2$ superconductors within both a toy two-band model and a more realistic
effective
low energy model including band structure specifics. We have found that,
 out of five types of pairing
states under consideration,
only  the $d_{x^2-y^2}$-wave pairing
gives rise to impurity resonance states.
The intra-gap states have energies far away from the Fermi energy, and they exist
regardless whether the inter-band scattering is present
or not. However, the inter-band scattering does tune the relative LDOS distribution of the resonance
states. Based on these results, we propose to use STM measurements of local electronic structure near nonmagnetic impurities to probe the pairing symmetry of the new superconductors.

One of the authors (J.-X.Z.) thanks J. Kunes, P. Wissgott, J. R. Yates, and Y.-S. Lee
for help with the WIEN2K interface and the Wannierization. J.-X.Z. also acknowledges
collaborations with T. Zhou, R. Beaird, I. Vekhter, and C. S. Ting for related research on iron-arsenic
superconductors.  A.V.B. acknowledges discussions with T. Das, Y. Bang, H. Ding,
M. Ogata, and H.-H. Wen. This work was supported by U.S. DOE  at
LANL  under Contract No. DE-AC52-06NA25396 and the DOE Office of Basic Energy
of Sciences (J.-X.Z. \& A.V. B.), and NSF Grant No. DMR-1006985 and the Robert A. Welch Foundation
Grant No. C-1411 (R.Y. \& Q.S.).

\end{document}